# The Ion Transition Range of Solar Wind Turbulence in the Inner Heliosphere: Parker Solar Probe Observations

S. Y. Huang[1], F. Sahraoui[2], N. Andrés[3,4], L. Z. Hadid[2], Z. G. Yuan[1], J. S. He[5], J. S. Zhao[6], S. Galtier[2], J. Zhang[1], X. H. Deng[7], K. Jiang[1], L. Yu[1], S. B. Xu[1], Q. Y. Xiong[1], Y. Y. Wei[1], T. Dudok de Wit[8], S. D. Bale[9], and J. C. Kasper[10]

[1]School of Electronic Information, Wuhan University, Wuhan, 430072, China

[2]Laboratoire de Physique des Plasmas, CNRS-Ecole Polytechnique-Sorbonne, Université Paris-Saclay, Observatoire de Paris-Meudon, Palaiseau, F-91128, France

[3]Instituto de Astronomía y Física del Espacio, CONICET-UBA, Ciudad Universitaria, 1428, Buenos Aires, Argentina

[4]Departamento de Física, Facultad de Ciencias Exactas y Naturales, UBA, Ciudad Universitaria, 1428, Buenos Aires, Argentina

[5]School of Earth and Space Sciences, Peking University, Beijing, 100871, China

[6]Purple Mountain Observatory, Chinese Academy of Sciences, 210023, Nanjing, China

[7]Insititute of Space Science and Technology, Nanchang University, Nanchang, 330031, China

[8]LPC2E, CNRS/CNES/University of Orléans, 45071 Orléans, France

[9]Space Sciences Laboratory and Physics Department, University of California, Berkeley, CA 94720-7450, USA

[10]Department of Climate and Space Sciences and Engineering, University of Michigan, Ann Arbor, MI 48109, USA

**Abstract**

The scaling of the turbulent spectra provides a key measurement that allows to discriminate between different theoretical predictions of turbulence. In the solar wind, this has driven a large number of studies dedicated to this issue using *in-situ* data from various orbiting spacecraft. While a semblance of consensus exists regarding the scaling in the MHD and dispersive ranges, the precise scaling in the transition range and the actual physical mechanisms that control it remain open questions. Using the high-resolution data in the inner heliosphere from Parker Solar Probe (PSP) mission, we find that the sub-ion scales (i.e., at the frequency $f \sim [2, 9]$ Hz) follow a power-law spectrum $f^\alpha$ with a spectral index $\alpha$ varying between -3 and -5.7. Our results also show that there is a trend toward and anti-correlation between the spectral slopes and the power amplitudes at the MHD scales, in agreement with previous studies: the higher the power amplitude the steeper the spectrum at sub-ion scales. A similar trend toward an anti-correlation between steep spectra and increasing normalized cross helicity is found, in agreement with previous theoretical predictions about the imbalanced solar wind. We discuss the ubiquitous nature of the ion transition range in solar wind turbulence in the inner heliosphere.

1. Introduction

Plasma, as the fourth state of matter in physics, is a gas consisting of roughly equal numbers of positively and negatively charged particles. It permeates various media such as the solar-terrestrial system, planetary magnetospheres, accretion disks, the interstellar medium and fusion devices (Baumjohann & Treumann, 1996). Turbulence, as a universal nonlinear phenomenon, plays a fundamental role in the mass transport and energy dissipation, which results in particle heating and/or acceleration in astrophysical plasmas (Tu & Marsch, 1995; Bruno & Carbone, 2005, 2013; Huang et al., 2012; Schekochihin et al., 2009; Lazarian et al., 2020; Sahraoui et al., 2020). Often, theoretical studies of hydrodynamic turbulence target the derivation of scaling laws that govern the turbulent fluctuations, which can be directly compared to the observations.

In the case of plasma turbulence, several popular analytical and numerical models have been developed to describe the turbulence cascade from the magnetohydrodynamic (MHD) to the electron scales, which led to different predictions on the scaling of the magnetic energy spectra (and other quantities, such as electric field spectra, velocity spectra etc). Those include the Iroshnikov-Kraichnan theory for isotropic MHD turbulence (with a $k^{-3/2}$ spectrum, Iroshnikov, 1963; Kraichnan, 1965), the anisotropic MHD turbulence and the so-called critical balance hypothesis ($k_\perp^{-5/3}$, Goldreich & Sridhar, 1995), the weak turbulence theory of Alfvén waves ($k_\perp^{-2}$, Galtier et al., 2000), the sub-ion scale cascade of whistler or kinetic Alfvén waves (KAW) turbulence (e.g., Biskamp et al. 1996; Li et al. 2001; Galtier & Bhattacharjee 2003; Galtier & Buchlin, 2007; Howes et al., 2008; Ghosh et al., 1996; Stawicki et al., 2001; David & Galtier, 2019), and an "ultimate" electron scale cascade (Schekochihin et al., 2009, Meyrand & Galtier, 2010; Camporeale & Burgess, 2011; Gary et al., 2012; Andrés et al., 2016a, 2016b; Zhao et al., 2013，2016), to name a few.

The solar wind plasma is an ideal natural plasma laboratory in which the above (and many other) theoretical predictions can be directly tested, given the large amount of *in-situ* data recorded by the early spacecraft such as Voyager, ACE, Wind, and Helios, and the more recent ones, e.g. Cluster, Magnetospheric MultiScale (MMS), and Parker Solar Probe (PSP). Usually, at MHD scales, the magnetic energy spectra generally exhibit an energy-containing range with a scaling $\sim f^{-1}$ for the frequencies $f < \sim 10^{-4}$ Hz (Bavassano et al. 1982; Bruno & Carbone, 2005; Kiyani et al., 2015; Sahraoui et al. 2020), and an inertial range with a scaling $\sim f^{-5/3}$ for the frequencies $\sim [10^{-4}, 10^{-1}]$ Hz (Coleman et al., 1968; Matthaeus & Goldstein, 1982; Goldstein et al., 1994; Leamon et al., 1998; Smith et al., 2006; Chen et al., 2019). In recent years, thanks to the high-time resolution data, the studies of turbulence have been focused more on the sub-ion scales (i.e., $f > 0.1$ Hz) where both dispersive and dissipative effects become important (Bale et al., 2005; Sahraoui et al., 2009, 2010, 2013; Kiyani et al., 2009; 2015; Alexandrova et al., 2012; Podesta et la., 2012; Salem et al., 2012; Chen et al., 2013; Wicks et al., 2013a; Andrés et al., 2014; Bruno et al., 2014; Duan et al., 2018; Huang et al., 2014, 2017; Goldstein et al., 2015; Huang & Sahraoui, 2019; Bowen et al., 2020a, 2020c). The spectra for the comprehensive "3D" (dispersion, dissipation, and diffusion) characteristics of space plasma turbulence between ion and electron scales have been provided (He et al., 2020). Near the ion spectral break point (typically observed between 0.1

Hz and 1 Hz), a steep ion-transition range (referred to as the dissipation range in early studies, see, e.g., Goldstein et al., 1994; Leamon et al., 1998; Smith et al., 2006) with a scaling up to $f^{-4.5}$ is sometimes observed (Sahraoui et al., 2010; Bruno et al., 2014; Kiyani et al., **2012**; He et al., 2015). This transition range has been interpreted as a consequence of an enhancement of the energy dissipation into ion heating by kinetic effects such as the Landau damping (e.g., Leamon et al., 1998; Sahraoui et al., 2010; Kobayashi et al., 2017) or the cyclotron damping (He et al., 2011; Woodham et al., 2018; Huang et al., 2020; Bowen et al., 2020c). A more recent theoretical model showed a possible competition between high imbalance and Landau damping (Miloshevich et al., 2020). Below the transition range ($f \sim [3, 30]$ Hz) a power-law ion dispersive range forms with an index that varies between -2.3 and -3.1 with a peak around -2.8 (Kiyani et al., 2013; Sahraoui et al., 2013). Besides, the steep transition-range spectra can also be produced by the weak turbulence of kinetic Alfven waves: spectra with power-law index -4 and even steeper (Voitenko & De Keyser, 2011).

As the electron scales are reached (i.e., $f > 40$ Hz), the magnetic spectra steepen again with spectral indices varying between -3.5 and -5.5 (e.g., Sahraoui et al., 2009, 2010, 2013). However, this power-law cascade scenario at sub-electron scales, although supported by most of the existing theoretical and numerical models, is challenged by some observations that suggested an exponential fall-off model (borrowed from neutral fluid turbulence), with a scaling $f^{-8/3}\exp(-f/f_d)$ from the ion to sub-electron scales (Alexandrova et al., 2012). Due to the low Signal-to-Noise (SNR) ratio of measurements in the solar wind at 1 Astronomical Unit (Sahraoui et al., 2013), the actual scaling of the magnetic energy spectra around electron scales remains an unsettled question. This limitation might be overcome by the data of the PSP mission taken in the next orbits closest to the sun.

In the present study, we use the high-time resolution data from the PSP mission in the inner heliosphere (around 0.17 AU) to investigate the scaling of the magnetic energy spectra from the MHD to sub-ion scales, typically $10^{-2}$ to $\sim 60$ Hz. Higher frequencies that would cover the electron scales are either not accessible to measurement in during the studied orbits because of the limited sampling frequency of the data ($f_N$ =293 Hz) or subject to high uncertainty because of the low-pass filtering applied on the data (Bowen et al., 2020b).

Besides confirming previous findings about the MHD scales, the distribution of the spectral indices from -3 to -5.7 at sub-ion scales clearly demonstrates the existence of a ubiquitous ion transition range in the inner heliosphere, while at 1AU its presence is only intermittently observed (Kobayashi et al., 2017). The paper is organized as follows: in section 2 we describe the observations data set and the selection criteria used in the present work. In section 3 we report our main observational results. Finally, in section 4 we provide a short discussion and the conclusion of this study.

## 2. Data Analysis

The recent NASA's PSP mission was launched in August, 12 2018 to investigate the processes of heating and acceleration of the solar wind and the solar energetic particles. Four instrument suites are carried by PSP to measure *in-situ* electric and magnetic fields, plasma and energetic particles, and to image the solar wind (Fox et al., 2016).

In the present study, we used the proton moments (density, velocity and temperature) with a sampling frequency of 1 Sa/cycle (1 cycle ≈ 0.873 s) for the Encounter mode provided by the Solar Wind Electron, Alpha, Proton (SWEAP) experiment (Kasper et al., 2016; Case et al., 2020); the magnetic field vector at the sampling frequency of 256 Sa/cycle (~ 292.969 Sa/s) measured by the FIELDS flux-gate magnetometer (FGM) and the search-coil magnetometer (SCM) for the Encounter mode (Bale et al., 2016; Malaspina et al., 2016; Jannet et al., 2020). In the present study, we analyzed observations from November 4$^{th}$ to November 7$^{th}$ 2018 during the first orbit of the PSP mission at a distance of about 0.17 AU, and used the merged FGM and SCM data to cover from MHD to electron scales (Bowen, et al. 2020b)

## 3. Observational Results

Figure 1 shows the observations from magnetic field, plasma velocity, plasma temperature, plasma density, plasma $\beta$ and the distance of PSP from the Sun. The magnetic field and proton velocity are presented in the heliospheric Radial-Tangential-Normal (RTN) coordinate system. The magnetic field shows large fluctuations (Figure 1a-1b) in the slow solar wind (i.e., $V_R$ < 500 km/s in Figure 3c) accompanied by large fluctuations in the proton density and temperature (Figure 1d-1e). The co-existence of kinetic Alfvén waves and Alfvén ion cyclotron waves was demonstrated in this slow solar wind during November 6$^{th}$ 2018 (Huang et al., 2020), similarly to the observations in the fast solar wind at 1 AU (e.g., He et al., 2011, 2012, 2015: Podesta, 2013; Bruno & Telloni, 2015). The plasma $\beta$ varies from 0.1 to ~10, but mostly less than 1 (Figure 1f).

To be able to study the magnetic energy spectra at the sub-ion scales for 15 mins windows, the Fourier transform was performed for the merged FGM and SCM measurements (Bowen et al., 2020b) to obtain the magnetic energy spectra for each 15 mins. The 15 mins duration ensures having at least one correlation time of the turbulent fluctuations (Parashar et al., 2020). Furthermore, to avoid possible "contamination" of the background turbulent spectra from intermittent but intense waves (e.g., ion cyclotron waves, whistlers), we excluded the magnetic spectra that showed clear spectral peaks from MHD to electron scales.

One should point out that the electron measurements are not available during this time interval. But we can estimate the Doppler shifted frequency of the electron inertial length $f_{de}$ ~ 180 Hz and the Doppler shifted frequency of the electron gyro-radius $f_{\rho e}$ ~ 320 Hz based on the assumptions of $N_e = N_i$ and $T_i/T_e = 5$. This indicates that the electron characteristic frequencies are larger than the Nyquist frequency of the magnetic field data provided by PSP, thus preventing us from investigating electron scale turbulence.

Figure 2a presents an example of the resulting magnetic spectra. Three distinct frequency bands can be clearly evidenced: i) the first one is inertial range with the slope of -1.71 below 1 Hz; ii) the second one is a steep ion-transition range ($f^{-3.63}$) at the scales between the proton cyclotron frequency and the Doppler shifted frequencies of the proton inertial length $d_i$ and gyro-radius $\rho_i$ (i.e., $f \sim [2, 9]$ Hz); iii) the third one has a scaling $f^{-2.68}$ between the ion and electron scales (~ [10, 60] Hz. These three distinct spectral features are also confirmed by all 219 selected spectra shown in Fig. 2b. However, the spectral indices have large variations in the ion transition range.

Figure 3 displays the histograms of the spectral indices at the MHD, sub-ion (transition range), and ion-to-electron scales, respectively. The slopes have a narrow distribution (from -1.95 to -1.35) at MHD scales (Figure 3a), consistent with the previous observations at 1 AU (e.g., Smith et al., 2006, **2011**), and a narrower distribution between ion and electron scales, i.e., from -2.78 to -2.37 (Figure 3c), highlighting the dominance of the scaling $f^{-2.63}$. However, the spectral indices have a broader distribution at sub-ion scales in the so-called transition range (from -2.75 to -5.76 in Figure 3b).

Figure 4a shows the correlation between the spectral indices at sub-ion scales and the integrated power of each spectrum near 0.1 Hz (i.e., from 0.1 Hz to 0.12 Hz). The corresponding power around 0.1 Hz is normalized to the minimum power $w_0$ around 0.1 Hz of all the magnetic spectra. One can see that the turbulent spectra essentially steepen at sub-ion scales for high values of the power near 0.1 Hz. The best fit is shown by the red curve that follows the equation: $slope = 12.2 - 15.9*(w/w_0)^{0.0092}$. Finally, the correlation coefficient between the spectral indices and the normalized power spectral densities is -0.2. Considering the scale-dependent of the normalized cross helicity in equation 10 in Wicks et al. (2013b), we calculated the mean normalized cross helicity over each time interval (i.e., time average) at different time lags (~ [9, 61] s), and found the correlation coefficients between the normalized cross helicity and the spectral indices is all below -0.3. Figure 4b displays the correlation between the spectral indices at sub-ion scales and the mean normalized cross helicity at time lag 25 s. The obtained values vary from 0.25 to 0.88. A vague anti-correlation can be seen between the spectra steepening at sub-ion scales and $\sigma_c$, with the best fit of *slope* = -2.1*$\sigma_c$ – 2.7. The correlation coefficient between the spectral indices and the cross helicity is -0.36.

In order to investigate the plasma environment in the presence of steep transition range at sub-ion scales in the inner heliosphere, Figure 5 presents the statistical histograms of the mean values of total magnetic field, plasma velocity, density, temperature, plasma *β* and the largest absolute correlation coefficient between magnetic field and plasma velocity for the intervals of all selected spectra. It can be seen that the total magnetic field can vary from 63 nT to 104 nT with peak around 86 nT, plasma velocity changes between 260 km/s to 480 km/s, plasma density is compact (100~460 $cm^{-3}$), plasma temperature varies from 20 eV to 88 eV with peak of 35 eV, and plasma *β* are mostly less than 1 with peak around 0.5, indicating that ion transition range could occur in different plasma environment. It is worth noticing that most of the values of the ***V-B*** correlation are larger than 0.8 (97.7%) with a peak around 0.92, implying that the transition range at sub-ion scales can be observed in essentially in highly Alfvénic, although a few cases (0.9%) seem to correspond to lower ***V-B*** correlation (~0.56), suggestive of less Alfvénic wind.

## 3. Discussion and Conclusions

The present study provides a statistical survey of the magnetic spectral features from the MHD to the sub-ion scales in the inner heliosphere using PSP measurements. Our results show that the magnetic energy spectral slopes have a low variability between the ion and electron scales (often referred to as the dispersive range, Stawicki et al., 2001), with the most probable index close to -2.63. The popular wisdom considers that this range is dominated by highly oblique KAW turbulence (Sahraoui et al., 2009; Salem et al., 2012; Chen et al., 2013) although a minor (quasi-parallel) component of the turbulence may consist of whistler waves

(or other type of fluctuations) (e.g., Podesta, 2013; Klein et al., 2014). This most probable index (-2.63) is slightly shallower than that ($\alpha \sim -2.8$) reported in early observations in the solar wind at 1AU and in the terrestrial magnetosheath (Sahraoui et al., 2013, 2020; Huang et al., 2014). This difference might be due to the compensation of the antialiasing filter implemented on the merged FGM and SCM data used in this work, which may alter the accuracy of the spectra at frequencies $\geq 40$ Hz (Bowen et al., 2020b).

On the contrary, at sub-ion scales our observations indicate a lack of universality of the turbulence scaling, with spectral slopes varying within the range [-3, -5.7], which are steeper than those reported in previous observations of solar wind turbulence at 1 AU (Smith et al., 2006; Kobayashi et al., 2017). This discrepancy remains to be explained.

The study revealed also the high variability of the spectral slopes at sub-ion scales and its dependence on the fluctuation amplitudes (or power) at the edge of the inertial range, which clearly points towards the lack of universality of the turbulent scaling near the ion scales, in agreement with the previous observations at 1 AU (Bruno et al., 2014; Kobayashi et al., 2017). In addition, it is found that the spectral slopes at sub-ion scales depend on the normalized cross helicity: the larger the cross helicity, the stepper the magnetic spectra. The cross helicity could illustrate the imbalance between the outward ($\delta z^+$) and inward ($\delta z^-$) propagating fluctuations (Vicks et al., 2013b): the larger the cross helicity, the more imbalanced the solar wind turbulence. Recent theoretical models showed the steep spectra can form at sub-ion scales as a consequence of high imbalance (Voitenko & De Keyser (2016)), or of a competition between high imbalance and Landau damping (Miloshevich et al., 2020). Our observations are consistent with those predictions.

Although several explanations have been proposed to account for the steep ion transition range, which include energy dissipation into ion heating by Landau or cyclotron damping (e.g., Leamon et al., 1998; Sahraoui et al., 2010; He et al., 2011; Woodham et al., 2018), weak turbulence of kinetic Alfven waves (Voitenko & De Keyser, 2011), dispersive turbulence and intermittency (e.g., Zhao et al., 2016) or a particular combination of waves and coherent structures (e.g., Lion et al., 2016), a unified picture of the physics occurring at the onset of the kinetic range remains to be built. This will be investigated in future.

**Acknowledgement**

This work was supported by the National Natural Science Foundation of China (41674161, 41874191, 41925018, 42074196), and the National Youth Talent Support Program. S.Y.H. appreciate the helpful discussions with Dr. Bowen. N.A. acknowledge financial support from the *Agencia de Promoción Científica y Tecnológica* (Argentina) through grant PICT 2018 1095. We thank the entire PSP team and instrument leads for data access and support. The SWEAP and FIELDS investigation and this publication are supported by the PSP mission under NASA contract NNN06AA01C. PSP data is publicly available from the NASA's Space Physics Data Facility (SPDF) at https://spdf.gsfc.nasa.gov/pub/data/psp/.

**Figure captions**

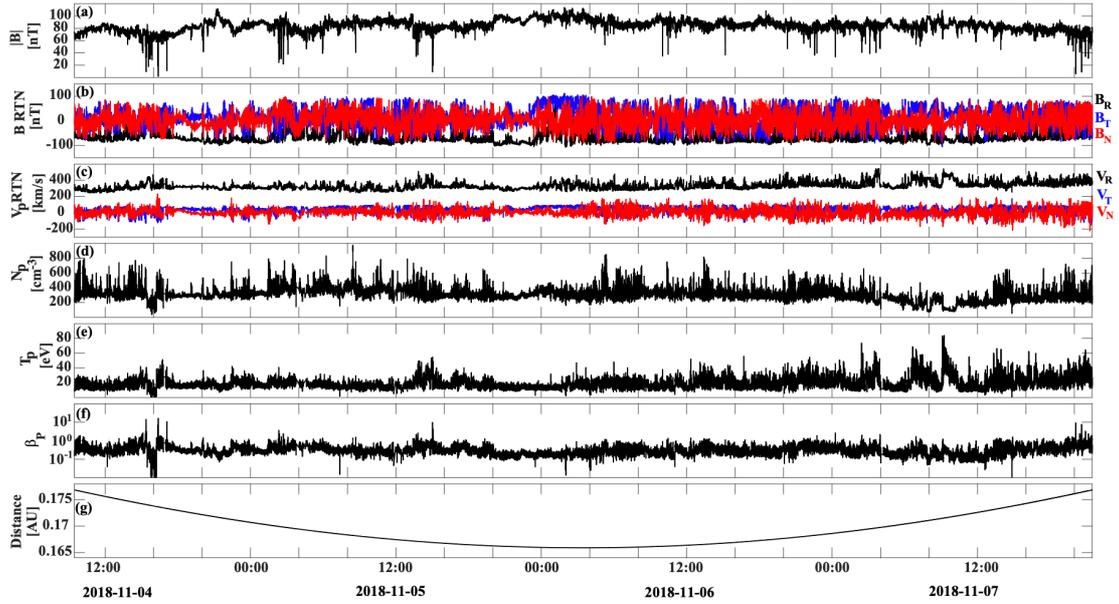

Figure 1. Overview observations from November 4th to November 6th, 2018. (a) The magnitude of magnetic field, (b) the three components of magnetic field (c) the three components of the proton velocity in RTN coordinates, (d) the proton number density, (e) the proton temperature, (f) the proton $\beta$, and (g) the distance away from the Sun.

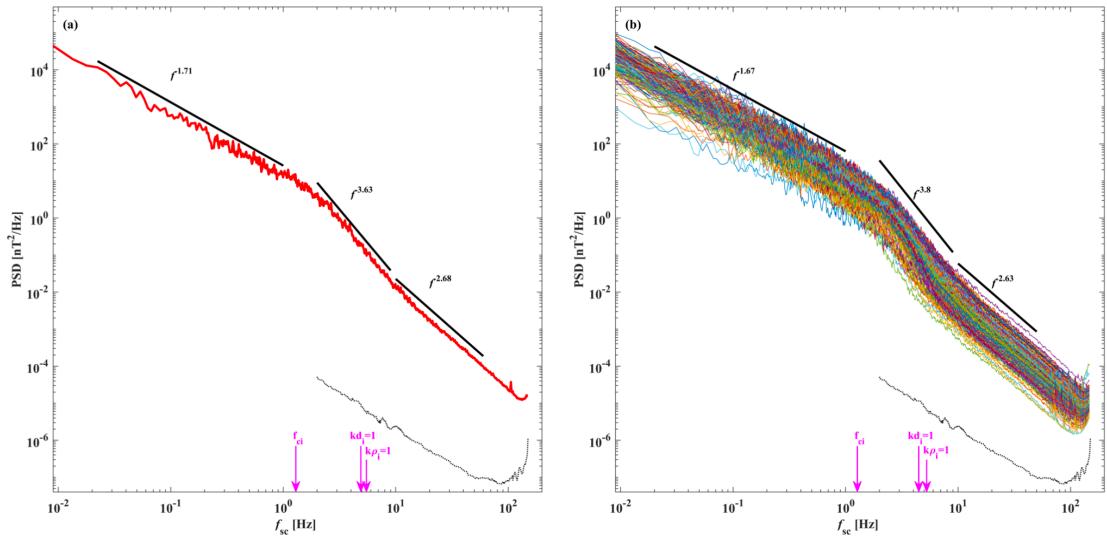

Figure 2. Power spectral densities of the magnetic field. (a) One example of magnetic spectrum during the time interval from 11:13:11 UT to 11:28:11 UT on November, 6 2018; (b) all selected spectra. The three vertical arrows from left to right correspond to the proton cyclotron frequency $f_{ci}$, the Doppler shifted frequency of the proton inertial length $kd_i=1$ and the proton gyroradius $k\rho_i=1$, respectively. The grey dotted curves display the noise of SCM instrument.

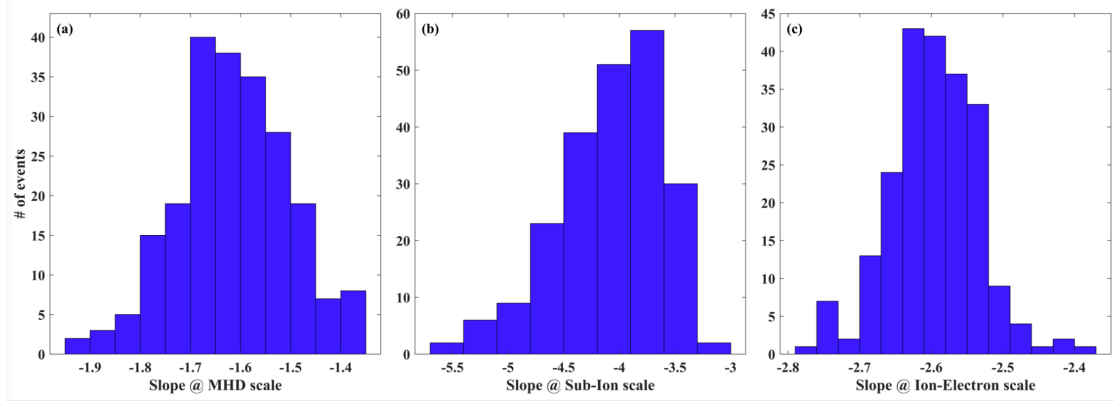

Figure 3. Histograms of spectral indices at three different ranges: (a) the MHD scales, (b) the sub-ion scales, and (c) the ion-electron scales.

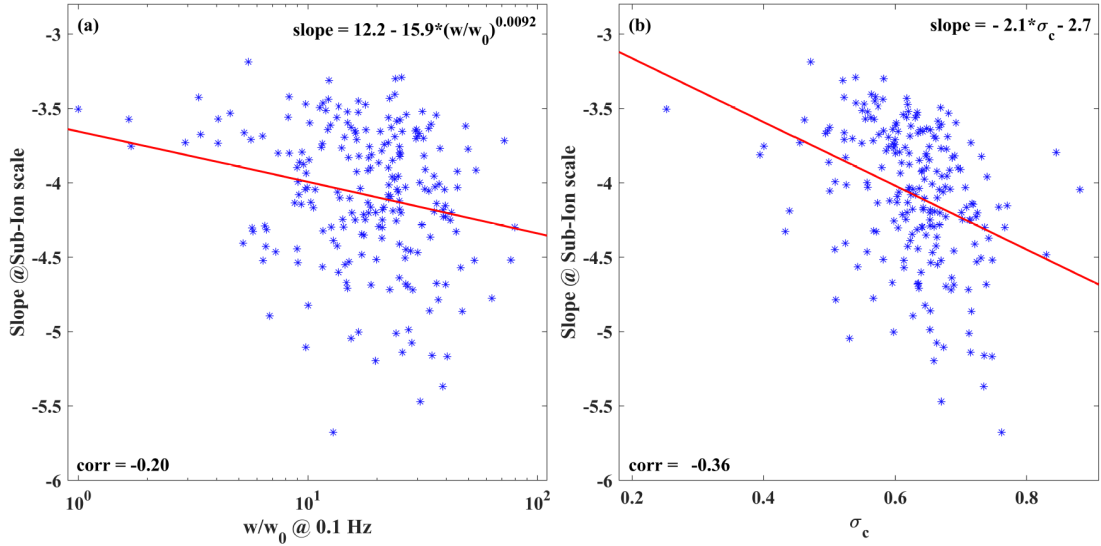

Figure 4. The spectral indices at sub-ion scales as a function of (a) the corresponding normalized power spectral density around 0.1 Hz ($w/w_0$) and (b) the normalized cross helicity. The $w_0$ is the minimum power spectral densities around 0.1 Hz for all selected magnetic spectra. The red line in (a) is the fitted curve given by $slope = 12.2 - 15.9*(w/w_0)^{0.0092}$. The cross helicity is average value during each time interval for all selected magnetic spectra. The red line in (b) is the fitted curve given by *slope* = -2.1*$\sigma_c$ – 2.7.

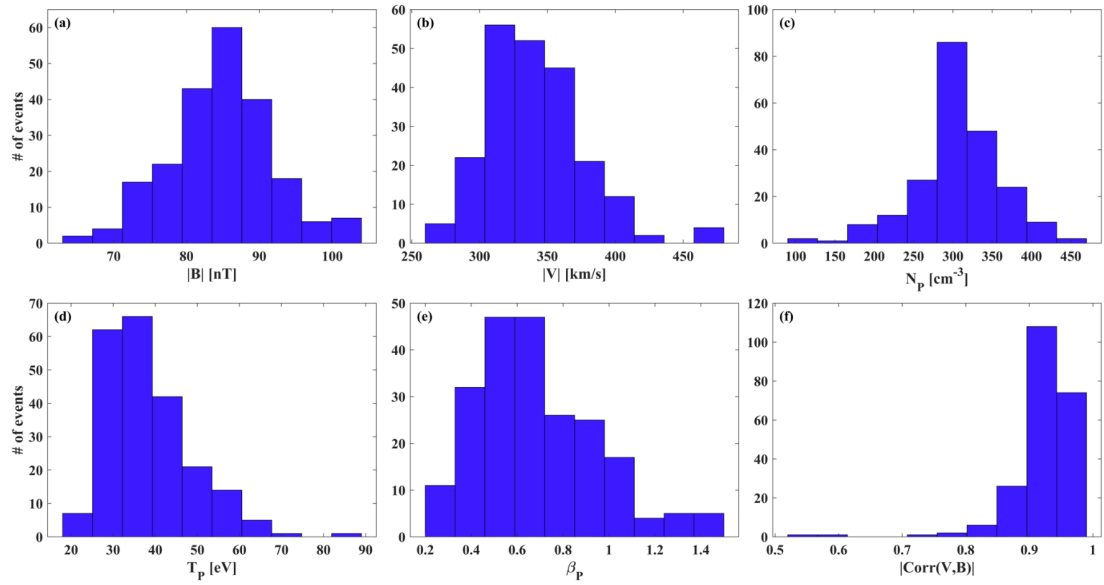

Figure 5. Histograms of the mean values of plasma parameters for all selected spectra: (a) the total magnetic field, (b) the total plasma velocity, (c) plasma density, (d) plasma temperature, (e) plasma *β*, and (f) the largest absolute values of the correlation coefficient between three components of magnetic field and the corresponding components of plasma velocity.